\documentclass{article}

\usepackage[margin=1in]{geometry}
\usepackage{authblk}
\usepackage{amssymb,verbatim,xcolor,fancyhdr,amsmath,amsthm}

\usepackage[utf8]{inputenc} 
\usepackage[T1]{fontenc}    
\usepackage{hyperref}       
\usepackage{url}            
\usepackage{booktabs}       
\usepackage{amsfonts}       
\usepackage{nicefrac}       
\usepackage{microtype}      
\usepackage{lipsum}
\usepackage{graphicx}
\graphicspath{ {./images/} }
\usepackage[english]{babel}
\newtheorem{theorem}{Theorem}
\newtheorem{lemma}{Lemma}
\newtheorem{remark}{Remark}
\newtheorem{corollary}[theorem]{Corollary}

\title{Improved and Generalized Algorithms for\\ Burning a Planar Point Set\thanks{This work is supported in part by the Natural Sciences and Engineering Research Council of Canada (NSERC). The work of the first author was supported by a MITACS Globalink Internship  at the University of Saskatchewan.}}

\author[1]{Prashant Gokhale}
\author[2]{J. Mark Keil} 
\author[2]{Debajyoti Mondal}

\affil[1]{Indian Institute of Science, Bangalore, India\\

  \texttt{prashantag@iisc.ac.in}}
\affil[2]{University of Saskatchewan, Saskatoon, Canada\\

  \texttt{\{mark.keil,d.mondal\}@usask.ca}}

\begin{document}
\maketitle  
\begin{abstract}
Given a set $P$ of points  in the plane, a point burning process is a discrete time process to burn all the points of $P$ where fires must be initiated at the given points. Specifically, the point burning process starts with a single burnt point from $P$, and at each subsequent step, burns all the points in the plane that are within one unit distance from the currently burnt points, as well as one other unburnt point of $P$ (if exists). The point burning number of $P$ is the smallest number of steps required to burn all the points of $P$. If we allow the fire to be initiated anywhere, then the burning process is called  an anywhere burning process, and the corresponding burning number is called anywhere burning number. Computing the point  and anywhere burning number is known to be NP-hard. In this paper we show that both these problems admit PTAS in one dimension. We then show that in two dimensions, point burning and anywhere burning   are $(1.96296+\varepsilon)$ and $(1.92188+\varepsilon)$  approximable, respectively, for every $\varepsilon 
>0$, which improves the previously known $(2+\varepsilon)$ factor for these problems. We also observe that a known result on set cover problem can be leveraged to obtain a 2-approximation for burning the maximum number of points in a given number of steps. We show how the results generalize if we allow the points to have different fire spreading rates.  
Finally, we prove that even if the burning sources are given as input, finding a point burning sequence itself is NP-hard. 

\end{abstract}


\graphicspath{ {./images/} }

\section{Introduction}
  
Graph burning was introduced by Bonato et al.~\cite{DBLP:conf/waw/BonatoJR14} as a simplified model to investigate the spread of influence in a network.  
Given a finite, simple, undirected graph $G$, the burning process on $G$ is defined as a discrete time process as follows. 
Initially, at $t=0$, all vertices in the graph are unburnt. Once a node is burnt, it remains so until the end of the process. At time $t = i$ ($i\geq 0$), the process burns all the neighbors of the currently burnt vertices, as well as one more unburnt vertex (if exists). This process stops when all vertices are burnt. The graph burning problem seeks to minimize the number of steps required to burn the whole graph.  We refer the reader to~\cite{DBLP:journals/cdm/Bonato21}  for a survey on graph burning.

\begin{figure}[pt]
    \centering
\includegraphics[width=\textwidth]{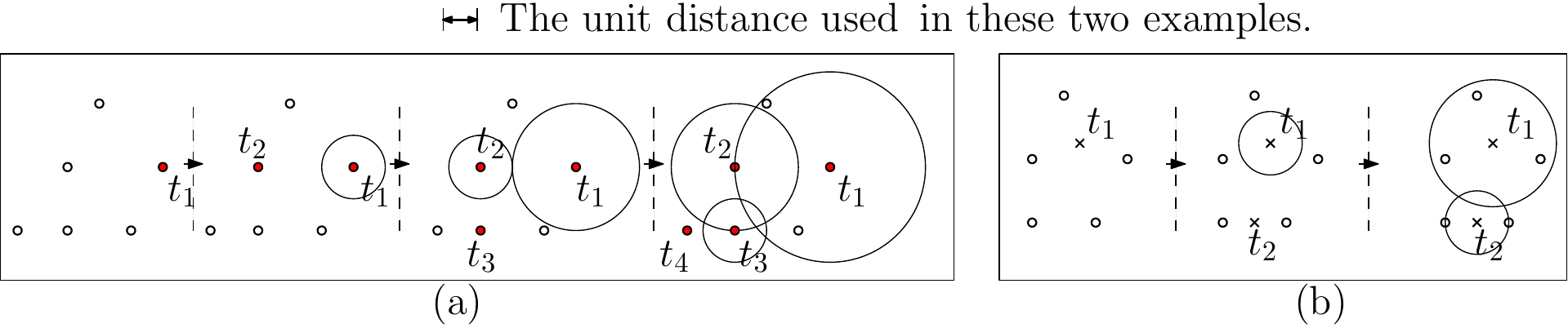}
    \caption{Illustration for (a) point burning and (b) anywhere burning. The burning sources are illustrated in labelled dots and cross marks, respectively. }
    \label{intro}
\end{figure}

Keil et al.~\cite{keil2022burning} introduced two geometric variants --- point burning and anywhere burning --- of this problem, where the goal is to burn a given set of points in the plane. The point burning model allows for initiating  fires only at the given points. The burning process starts by burning one given point, and then at each subsequent step, the fire burns  all unburnt points of the plane that are within one unit of any currently burnt point  and a new unburnt given point is chosen to initiate the fire. Figure~\ref{intro}(a) illustrates this model.  In the anywhere burning model, the burning process is the same but the fires can be started at any point in the plane. Figure~\ref{intro}(b) illustrates this model.  Note that we may not have an unburnt vertex to initiate fire at the last step. 

The geometric version of the burning process can provide a simple model of supply management where products need to be shipped in bulk to distribution centers. Consider a business that needs to maintain a continuous supply of perishable goods to a set of $P$ locations. Each day it can manage to send one large shipment to a hub location that distributes the goods further to the nearby locations over time. The point burning considers only the points of $P$ as potential hubs, whereas anywhere burning allows to create a hub at any point in the plane. The burning number indicates the minimum number of days needed to distribute the goods to all locations. For example, in Figure~\ref{intro}(a), the hubs are $t_1$, $t_2$, $t_3$ and $t_4$, and the business can keep sending the shipments to the hubs after every three days in the same order.

The graph burning problem is known to be NP-hard for forest of paths~\cite{Bessy2017} and  APX-hard for subcubic graphs~\cite{DBLP:conf/walcom/MondalP0R21}. However, the problem is  approximable within a factor of 3~\cite{bonato2019approximation}, which has recently been improved further to $(3-2/b)$ where  $b$ is the burning number of the input graph~\cite{DBLP:journals/access/Garcia-DiazSRC22}. The introduction of point and anywhere burning naturally raises the question of whether one can prove analogous results for these problems. Keil et al.~\cite{keil2022burning} showed that both problems are NP-hard, but approximable  within a factor of $(2+\varepsilon)$, for every $\varepsilon >0$. However, a number of interesting problems are yet to be explored. For example, can we find better approximation algorithms? Does there exist a PTAS for these problems? Can we maximize the number of burnt points within a given time limit? What happens if the points have different rates for spreading the fire? This is relevant in practice when the distribution capabilities vary across different  distribution centers. Can we find a burning sequence in polynomial time if the burning sources are given? This last question is known to be NP-complete for graph burning~\cite{DBLP:conf/walcom/MondalP0R21}.


\smallskip
\noindent
\textbf{Contribution:} In this paper, we obtain  the following results.
\begin{itemize}
    \item We show that in one dimension one can find a PTAS for both point and  anywhere  burning. In two dimensions, we improve the approximation ratio for point and anywhere burning to $(1.96296+\varepsilon)$ and $(1.92188+\varepsilon)$, respectively. 
    
    \item We consider a generalization where the fire spreading rates vary across the  given  points. We show how to adapt the existing approximation algorithms to obtain constant-factor approximation for point burning if the ratio of the largest and the smallest rate is a constant. 
    
    \item We prove that even if the burning sources are given as input, finding a point burning sequence itself is NP-hard. This problem was known to be NP-hard for graph burning, but the same hardness reduction does not hold in geometric setting. 
    
    \item Our NP-hardness result implies that given a set of $q$ burning sources, it is NP-hard to find a point burning sequence that maximizes the number of burnt  points within $q$ steps. In contrast, we show how to adapt a set cover technique to obtain a 2-approximation for burning the maximum number of points in a give number of steps. This result holds even when a set of burning sources are specified at the input.
\end{itemize}

\section{Burning Number in One Dimension}

In this section we consider the case when the points of $P$ are on a line. Assume that the points are ordered in increasing x-coordinate and let $A[i]$ be the x-coordinate of the $i$th point from the left. Let $\delta^{*}$ be the burning number. 

\subsection{PTAS for Anywhere Burning}

We now provide a polynomial time approximation scheme (PTAS), i.e., a $(1+\varepsilon)$-approximation algorithm for every $\varepsilon>0$, for the anywhere burning problem. Intuitively, we can visualize this problem as a covering problem with intervals in one dimension.

Our strategy is to make a guess $\delta$ for the burning number starting from $1$. We keep increasing the guess by 1 as long as we can prove the current $\delta$ to be a lower bound on the burning number. At some point when we are unable to establish $\delta$ as a lower bound, we  show how to find an approximate solution.

Note that for a $\delta$, we have $\delta$ intervals of length  $0$, $2$, $4$, ... , $2\left(i-1\right)$ to burn all the points. We group these intervals into $t$ different groups as follows. The first group will have $\frac{\delta}{t}$ intervals with length at most $2\left ( \frac{\delta}{t}\right)$. Generally, the $j^{th}$ group will have $\frac{\delta}{t}$ intervals with length larger than $2(j-1)\left ( \frac{\delta}{t}\right)$ and at most $2j\left ( \frac{\delta}{t}\right)$. For each group, we now relax all its intervals such that their length is equal to the largest interval in the group, i.e., $2j\left ( \frac{\delta}{t}\right)$. We use the notation $S(\delta)$ to denote this new set of intervals. We now use dynamic programming to check if there is a placement of the intervals in $S(\delta)$ so that every point is covered. If not, then we are sure that $\delta$ is not the burning number, as we had relaxed every interval. Otherwise, we will use these intervals to obtain an approximate solution. Later, we will show how to choose  $t$ to obtain a PTAS.

Let $V=(v_1,\ldots,v_{t+1})$ be a $(t+1)$-tuple of integers. Let $D(V)$ be the problem of covering $v_{t+1}$ points of $P$ from the left with $v_j$ intervals of group $j$, where $1\le j\le t$. We use $P(v_{t+1})$ to denote these points that are to be covered. Assume that the rest of the points, i.e., $P\setminus P(v_{t+1})$ are already covered by the rest of the intervals of $S(\delta)$. We can then express $D(V)$ using the following recursion with trivial values for the base cases:
$D(V) = \bigvee_{j=1}^{t} D(W^j).$

Here $W^j=(w_1,\ldots,w_{t+1})$ is similar to $V$ except at two places: $w_j$ and $w_{t+1}$. The $j$th element $w_j$ is set to $(v_j-1)$, because we have now used  one more interval $I$ of group $j$ to cover   some points of $P(v_{t+1})$. Note that the best position for this interval is when its right end coincides with the rightmost point $z$ of $P(v_{t+1})$. It is straightforward to prove this formally with the observation that for every covering we can shift the rightmost interval to the left unless its right end point coincides with $z$. Therefore, we set $w_{t+1}$ to be the number of remaining points that remains to be covered after placing $I$.

We store the solution to the subproblems using a multidimensional table $L$ where $L[V]$ stores the solution to $D(V)$. Using table lookup, the time taken to compute a cell of the table is $O(t)$. Since each of the $t$ groups has $\frac{\delta}{t}$ intervals, and since $P$ has $n$ points, the size of $L$ is $O\left(n\left(\frac{\delta}{t}\right)^{t}\right)$. 
Thus  the overall running time of the dynamic programming algorithm is  $O\left(nt\left(\frac{\delta}{t}\right)^{t}\right)$.

Note that the question of covering $P$ using $S(\delta)$ is obtained from $L[U^{\delta}]$, where $U^{\delta}=(u_1,\ldots,u_{t+1})$ be a $(t+1)$-tuple with $u_j = \frac{\delta}{t}$ for $1 \leq j \leq t$ and $u_{t+1}  = n$. We now describe the process of guessing $\delta$ and arriving at an approximate answer.

\begin{itemize}
    \item Start guessing from $\delta = 1$
    \item Use the dynamic programming to compute $L[U^{\delta}]$, i.e., the solution to  the relaxed covering question for $S(\delta)$  in $O\left(nt\left(\frac{\delta}{t}\right)^{t}\right)$ time.
    \item If $L[U^{\delta}]$ does not contain an affirmative answer, then we know that $\delta^* > \delta$. We thus  iterate again by increasing value of $\delta$ by $1$.  
    \item If $L[U^{\delta}]$  contains an affirmative answer, stop and return the approximate burning number to be $\delta \left( 1 + \frac{2}{t} \right)$. At this point we know that $\delta \leq \delta^{*}$. We construct the burning sequence as follows: First, burn the midpoints of all intervals of length $2\delta$ in the covering solution (i.e., the largest intervals), then burn the midpoints of all intervals of length $2\left(t-1\right)\left ( \frac{\delta}{t}\right)$ in the solution and so on. However, since we used relaxed intervals in the dynamic programming, we keep burning for an extra $\frac{2\delta}{t}$ steps to ensure that each  interval reaches its relaxed size (i.e., burns all points of $P$). 
\end{itemize}

Observe that we took $\delta + \frac{2\delta}{t} = \delta \left( 1 + \frac{2}{t} \right)$ steps to burn all the points. 
Since $\delta\le \delta^*$, we have 
$
\delta\left( 1+\frac{2}{t} \right) \leq \delta^*\left( 1+\frac{2}{t} \right)
$. Therefore, our algorithm achieves an approximation factor of $1+\frac{2}{t}$ for anywhere burning. Since $\delta^{*} \leq n$, the  total running time of our algorithm is bounded by $O\left(n^2\left(\frac{n}{t}\right)^{t}\right)$. 
Given an $\varepsilon>0$, we choose $t$ such that $\varepsilon = \frac{2}{t}$. Thus we get an approximation factor of $(1+\varepsilon)$ and a running time of  $O\left(n^{2+\frac{2}{\varepsilon}} (\frac{\varepsilon}{2})^{\frac{2}{\varepsilon}}\right)$. We thus obtain the following theorem.



\begin{theorem}
Given a set $P$ of $n$ points on a line and a positive constant $\varepsilon>0$. One can approximate the anywhere burning number of $P$ within a factor of $(1+\varepsilon)$ and compute the corresponding burning sequence in polynomial time.  
\end{theorem}

\subsection{PTAS for Point Burning}

We can slightly modify the algorithm for anywhere burning problem to obtain a PTAS for the point burning problem.  

The strategy is the same as guessing the burning number $\delta$ from 1 to $n$. Similar to anywhere burning, for each guess $\delta$, we use a dynamic programming algorithm to check whether we can establish $\delta$ as a lower bound. If so, then we increase $\delta$, otherwise, we construct an approximate solution. 

We define $S(\delta)$  in the same way as we did for anywhere burning. Since the point burning restricts the burning sources to be at the given points, we modify the dynamic programming to check whether the set $S(\delta)$ can cover $P$ by placing the midpoints of the intervals at some of the given points. The corresponding recurrence relation is as follows.

\begin{displaymath}
D(V) = \bigvee_{j=1}^{t} D(W^j),
\end{displaymath}
Here $W^j=(w_1,\ldots,w_{t+1})$ is similar to $V$ except at two places: $w_j$ and $w_{t+1}$. The $j$th element $w_j$ is set to $(v_j-1)$, because we have now used an  interval $I$ of group $j$ to cover   some points of $P(v_{t+1})$. Let $z$ be the rightmost point of $P(v_{t+1})$. Then the best position for $I$ is determined by a point $q$ of $P(v_{t+1})$ such that placing the midpoint of $I$ at any point to the left of $q$ fails to cover $z$. We thus place $I$ such that its midpoint coincides with $q$ and set $w_{t+1}$ to be the number of remaining points that remains to be covered after placing $I$.

 The rest of the algorithm is the same as for anywhere burning. 

\begin{theorem}
Given a set $P$ of $n$ points on a line and a positive constant $\varepsilon>0$. One can approximate the point burning number of $P$ within a factor of $(1+\varepsilon)$ and compute the corresponding burning sequence in polynomial time.  
\end{theorem}



\section{Burning Number in Two Dimensions}
In this section we assume that the points of $P$ are in $\mathbb{R}^2$. We first give a $(1.92188+\varepsilon)$-approximation algorithm for anywhere burning (Section~\ref{sec:ab}) and then a $(1.96296+\varepsilon)$-approximation algorithm for point burning (Section~\ref{sec:pb}). Note that this improves the previously known $(2+\varepsilon)$-approximation factor for these problems~\cite{keil2022burning}.

\subsection{Anywhere Burning}
\label{sec:ab}
Our algorithm for anywhere burning is inspired by the $(2+\varepsilon)$-approximation algorithm of Keil et al.~\cite{keil2022burning}, where we improve the approximation factor by using a geometric covering argument. 


Keil et al.~\cite{keil2022burning} leverage the discrete unit disk cover problem to obtain an approximation algorithm for anywhere burning. The input to the \emph{discrete unit disk cover problem} is a set of points $P$ and a set of unit disks $\mathcal{U}$ in $\mathbb{R}^2$. The goal is to choose the smallest set $U \subset \mathcal{U}$ that covers all the points of $P$. There exists a PTAS for the discrete
unit disk cover problem~\cite{DBLP:journals/dcg/MustafaR10}. 

Let $\delta^{*}$ be the actual burning number. Keil et al.~\cite{keil2022burning} iteratively guess the anywhere burning number $\delta$ from $1$ to $n$. For each $\delta$, they construct a set of $n$ disks, each of radius $\delta$, that are centered at the points of $P$, and $\left(n \choose 3\right) + \left(n \choose 2\right)$ additional  disks, where each disk is of radius $\delta$ and is centered at the center of a circle determined by either two or three points of $P$. The reason is that any solution to the anywhere burning can be perturbed to obtain a subset of the discretized disks. Then they compute a $(1+\varepsilon)$ approximation $U_{\delta}^{'}$ for the discrete unit disk cover $U_{\delta}$. If $\frac{|U_{\delta}^{'}|}{(1+\varepsilon)} > \delta$, then $\delta$ cannot be the burning number as otherwise, one could construct a smaller discrete unit disk cover by choosing disks that are centered at the burning sources of an optimal burning sequence. At this point, the guess is  increased by one. The iteration stops when  $\frac{|U_{\delta}^{'}|}{(1+\varepsilon)} \leq \delta$, where we know that $\delta \leq \delta^{*}$. At this point, Keil et al.~\cite{keil2022burning} show how to construct a burning sequence of length $(2+\varepsilon)$. 

We now describe a new technique for constructing the burning sequence. 
To burn all points in $P$, we use $1.92188 \delta(1+\varepsilon)$ steps. We first choose the centers of  a $0.92188$ fraction of $U_{\delta}^{'}$ disks and   burn them in arbitrary order. This requires $0.92188|U_{\delta}^{'}| = 0.92188 \delta(1+\varepsilon)$  steps. We then burn for another  $\delta(1+\varepsilon)$ steps. This will ensure the previously chosen $0.92188 |U_{\delta}^{'}|$  fires to  have a radius of at least $\delta$. Therefore, these fires will burn all the points that are covered by the corresponding disks of the discrete unit disk cover solution.  We are now left with $(1-0.92188)|U_{\delta}^{'}|=0.07812|U_{\delta}^{'}|$ disks  in $U_{\delta}^{'}$ that   need to be covered using the next $\delta(1+\varepsilon)$ steps.  Observe that $(1-0.6094)=0.3906$ of these   $\delta(1+\varepsilon)$ fires  have radius at least $0.6094$.  Since a unit disk can be covered by 5 equal disks\footnote{\url{http://oeis.org/A133077}} of radius at most 0.6094~\cite{neville1915solution}, we can cover the remaining  $\frac{0.3906}{5}|U_{\delta}^{'}|= 0.07812|U_{\delta}^{'}|$ disks of $U_{\delta}^{'}$.


 
 Since $\delta \le \delta^*$, We have the following theorem.

 

\begin{theorem}\label{th:anywhere}
Given a set $P$ of points in $\mathbb{R}^2$ and an $\varepsilon>0$, one can compute an anywhere burning sequence in polynomial time where the length of the sequence is at most $1.92188(1+\varepsilon)$ times the anywhere burning number of $P$.
\end{theorem}

\subsection{Point Burning}
\label{sec:pb}
The algorithm for anywhere burning can be easily adapted to provide a $(\frac{53}{27}+\varepsilon)$ approximation ratio for point burning. We will use the following geometric configuration. 

 \begin{figure}[pt]
     \centering
 \includegraphics[width=.45\textwidth] 
 {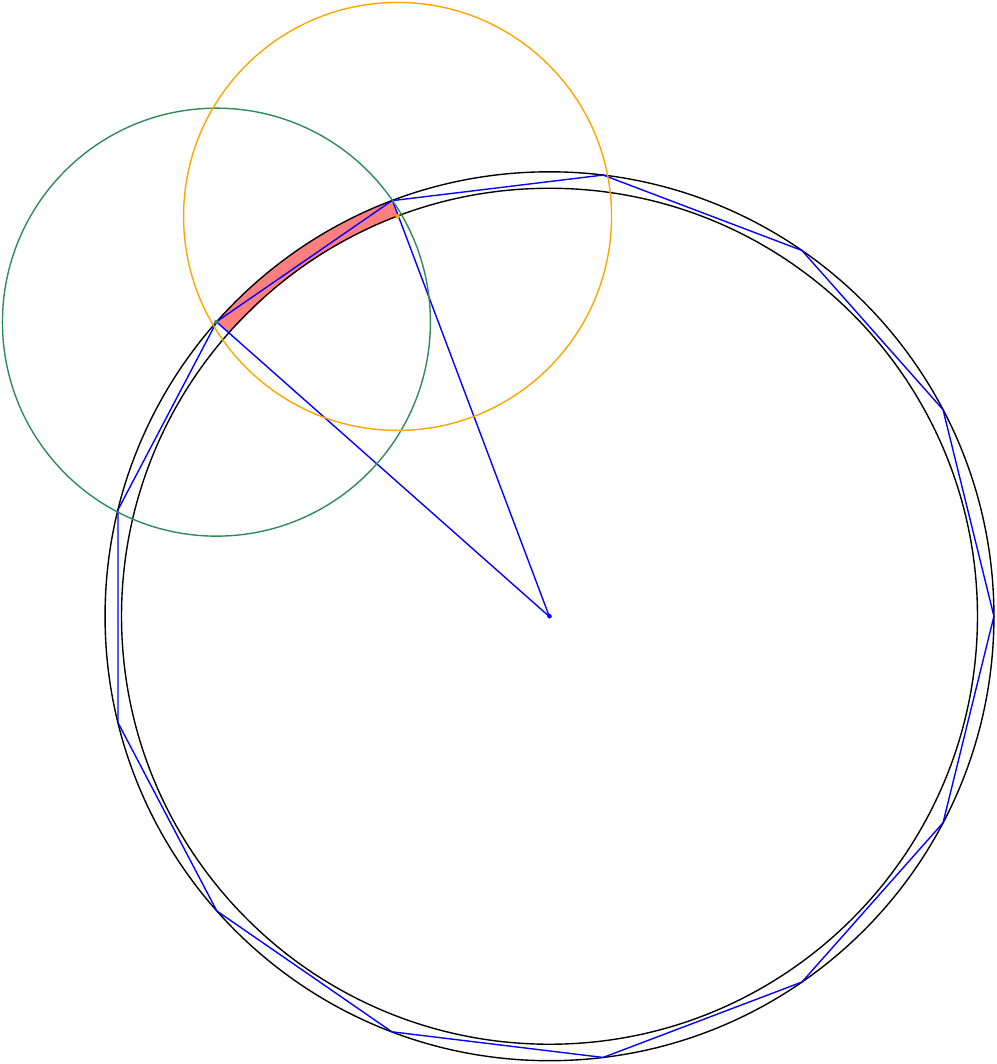}
     \caption{Illustration for Remark~\ref{lem:ann2}. A zone is highlighted in red shaded region. Each of the green and orange circles is of radius $\frac{13}{27}$ and covers the entire zone.}
     \label{ann2}
 \end{figure}
\begin{remark}\label{lem:ann2}
Consider the annulus  defined by two circles of radii $r_1 = 1$ and $r_2 = \frac{26}{27}$.  Let $C_1 $ be the circle with radius $r_1$ and let $Q_1$ be a 13 sided regular-polygon inscribed inside $C_1$. Split the annulus into 13 equal area regions (zones) by drawing lines from the center of the annulus to the corners of the polygon. Then every circle with center inside a zone and with radius $\frac{13}{27}$ covers that entire zone. Figure~\ref{ann2} illustrates this configuration.
\end{remark}
\begin{proof}
The claim is straightforward to verify by measuring the pairwise distances of the four corners of a zone.
\end{proof}

Similar to anywhere burning, for a guess $\delta$, we construct an instance of a discrete unit disk cover problem. However, here we use  a set of $n$ disks, each of radius $\delta$, centered at the points of $P$. Then we follow the same guessing strategy and stop as soon as we have  $\frac{U_{\delta}^{'}}{(1+\varepsilon)} \leq \delta$. We then first burn all the points corresponding to $U_{\delta}^{'}$, and burn for an additional $\frac{26 \delta(1+\varepsilon)}{27}$ steps. We are now left with $\frac{\delta(1+\varepsilon)}{27}$ annuli which we need to cover in these additional steps. We divide each annuli into thirteen zones, as in Remark~\ref{lem:ann2}. 
We chose at most one point of $P$ in each zone  (if exists), and burn that point during the first $\frac{13 \delta(1+\varepsilon)}{27}$ extra steps. Observe that there can be at most $\frac{13 \delta(1+\varepsilon)}{27}$ such points over all annuli. Furthermore, the fire around each of these points will reach at least  a radius of $\frac{13 \delta(1+\varepsilon)}{27}$ and by Remark~\ref{lem:ann2}, it will engulf its corresponding zone completely. Therefore,  all points will be burnt within  $\frac{53(1+\varepsilon)}{27}  \approx 1.96(1+\varepsilon)$ steps.

\begin{theorem}\label{th:anywhere}
Given a set $P$ of points in $\mathbb{R}^2$ and an $\varepsilon>0$, one can compute a point  burning sequence in polynomial time where the length of the sequence is at most $\frac{53(1+\varepsilon)}{27}\approx 1.96(1+\varepsilon)$ times the point burning number of $P$.
\end{theorem} 

\section{Generalizations for Point Burning}

In this section we consider two generalizations of the problem. 

The first one is \emph{point burning with non-uniform rates}  (Section~\ref{nonuniform}). Specifically, for each $i$ from 1 to $n$, the $i$th point in $P$ is assigned a positive integer (\emph{rate}) $r_i$. If a  fire starts at the $i$th point, the fire will spread with a rate of $r_i$ per step. Note that point burning with uniform rates, i.e., $r_1=\ldots=r_n$ reduces to point burning. 
The second one is \emph{$k$-burning number}, i.e., when $k$ points can be burned at each step (Section~\ref{kburn}). This version has previously been considered for graph burning and graph $k$-burning number is known to be $3$ approximable~\cite{DBLP:conf/walcom/MondalP0R21}.  

\subsection{Point Burning with Non-uniform Rates}\label{nonuniform}

Let $h$ be the ratio of the fastest rate to the slowest rate, i.e., 
$h = \max_{1 \leq i,j \leq n} \frac{r_i}{r_j}$ (intuitively, it is the maximum ratio over all pairs of rates). In this section we show that for every fixed $h$, point burning number with non-uniform rates  is approximable within a constant factor. 
We slightly modify our point burning algorithm described in Section~\ref{sec:pb} by leveraging a dominating set in a disk graph, as follows.  A \emph{disk graph} is a  geometric intersection graph where the vertices correspond to a set of disks in the plane and there is an edge if and only if the corresponding pair of  disks intersect. A  \emph{dominating set} in a disk graph is a subset $S$ of vertices such that every vertex is either in $S$ or has a neighbour in $S$. 
 Gibson and Parwani~\cite{gibson2010algorithms} provides a PTAS for finding a  minimum dominating set in disk graphs, where the disks can have different radii. 

We are now ready to present the algorithm. 
For a positive integer $m$, let $G_m$ be the disk  graph obtained by constructing for each point $t \in P$, a disk centered at $t$ with radius   $\frac{m}{2} r_t$. Let $D_m$ be a minimum dominating set of $G_m$. 


We  start guessing the burning number $\delta$  from $1$ to $n$, and for each guess, we compute a $(1+\varepsilon)$-approximate dominating set $E_{\delta - 1}$ of $G_{\delta - 1}$. We now have $|E_{\delta - 1}| \leq (1 + \varepsilon)|D_{\delta - 1}|$. If $\delta < \frac{|E_{\delta - 1}|}{(1+\varepsilon)} \leq |D_{\delta - 1}|$, then we can claim that $\delta$ burning sources are not enough to burn all the points and can increase the guess by 1. Suppose for a contradiction that all the points can be burned in $\delta$ steps. We can then choose the disks corresponding to the  burning sources to obtain a dominating set with less than $|D_{\delta - 1}|$ disks, a contradiction.

Once we get $\delta \geq \frac{|E_{\delta - 1}|}{(1+\varepsilon)}$, we stop. At this point, we know that $\delta \leq \delta^*$. We now construct a burning sequence by first burning  all the points in $E_{\delta - 1}$ (in an arbitrary order) and then continuing the   burning for $h(\delta - 1)$ more steps. We need to show all the points of $P$ are burned. Take some point $p \in P$, if $p \in E_{\delta - 1}$, then it is clearly burned. Otherwise, $p$ is dominated by a point $q \in E_{\delta - 1}$. Here the Euclidean distance between $p$ and $q$ is at most $\frac{\delta - 1}{2} (r_p + r_q)$. 

If $r_q \geq r_p$, then the radius for the fire initiated at $q$ is $r_q(\delta -1)$ which is larger than  $\frac{\delta - 1}{2} (r_p + r_q)$. Therefore, $p$ must be burned.

Otherwise, assume that  $r_q < r_p$. Here the distance between $p$ and $q$ is at most $\frac{\delta - 1}{2} (r_p + r_q) \leq  (\delta - 1) r_p$. Since $q\in E_{\delta - 1}$, by our burning strategy,  the fire at $q$ will continue to burn for at least $h(\delta - 1)$ steps. Therefore, its radius is at least  $r_qh(\delta - 1)$ steps. Since $h$ is the maximum ratio of the burning rates, $r_qh(\delta - 1) \geq r_q (\frac{r_p}{r_q})  (\delta - 1) = (\delta - 1) r_p$. Hence the point $p$ must be burned. 

The number of rounds taken by our algorithm is  $|E_{\delta - 1}| + h(\delta - 1) \leq (1 + \varepsilon) \delta^* + h(\delta^* - 1) = (1+h+\varepsilon) \delta^*$. We thus obtain the following theorem.

\begin{theorem}\label{th:nu}
Let $P$ be a set of points in $\mathbb{R}^2$, where each point is assigned a burning rate.   Let $h$ be the ratio of the fastest rate to the slowest rate. Given an $\varepsilon>0$, one can compute a point burning sequence in polynomial time where the sequence length  is at most $(1+h+\varepsilon)$ times the point burning number of $P$.
\end{theorem} 

\subsection{$k$-Burning with Non-uniform  Rates}
\label{kburn}

We now consider the point burning model when $k$ points are allowed to burn at each step and the goal is to compute the $k$-burning number, i.e., minimum number of rounds to burn all points of $P$.  
Our algorithm for this model is the same as in the previous section except that we stop   iterating the guess  when  $k\delta \geq \frac{|E_{\delta - 1}|}{(1+\varepsilon)}$. The reason we keep iterating in the case when $k\delta < \frac{|E_{\delta - 1}|}{(1+\varepsilon)}<|D_{\delta - 1}|$ is that   burning all points in $k\delta$ steps would imply the existence of a dominating set of size smaller than $|D_{\delta - 1}|$. We thus obtain the following theorem.

\begin{theorem}\label{th:nukb}
Let $P$ be a set of points in $\mathbb{R}^2$, where each point is assigned a burning rate, and let $h$ be the ratio of the fastest rate to the slowest rate. Given an $\varepsilon>0$ and a positive integer $k>0$, one can compute a point $k$-burning sequence in polynomial time where the length of the sequence is at most $(1+h+\varepsilon)$ times the point burning number of $P$.
\end{theorem} 

\section{NP-Hardness} 
In this section we show that computing a point burning sequence is NP-hard even if we are given the burning sources. 

We will reduce the NP-Hard problem LSAT~\cite{arkin2018selecting}, which is a 3-SAT formula where  each clause (viewed as a set of literals) intersects at most one other clause, and, moreover, if two clauses intersect, then they have exactly one literal in common. Given an LSAT instance, one can sort the literals such that each clause corresponds to at most three consecutive literals, and each clause may share at most one of its literals with another clause, in which case this literal is extreme in both clauses~\cite{arkin2018selecting}.

Let $I$ be an instance of LSAT with  $m$ clauses and $n$ variables. Without loss of generality we may assume for every variable, both its positive and negative literals appear in $I$. Otherwise, we could set a truth value to the variable to satisfy and eliminate some clauses to obtain an LSAT instance $I'$ which is satisfiable if and only if $I$ is satisfiable.  We will construct a point set $P$ with  $4n+m$ points and identify $2n$ points to be used as the burning sources. We will show that $I$ is satisfiable if and only if $P$ has a point burning sequence that uses only the given $2n$ points as the burning sources. 

\begin{figure}[pt]
    \centering
    \includegraphics[width=\textwidth]{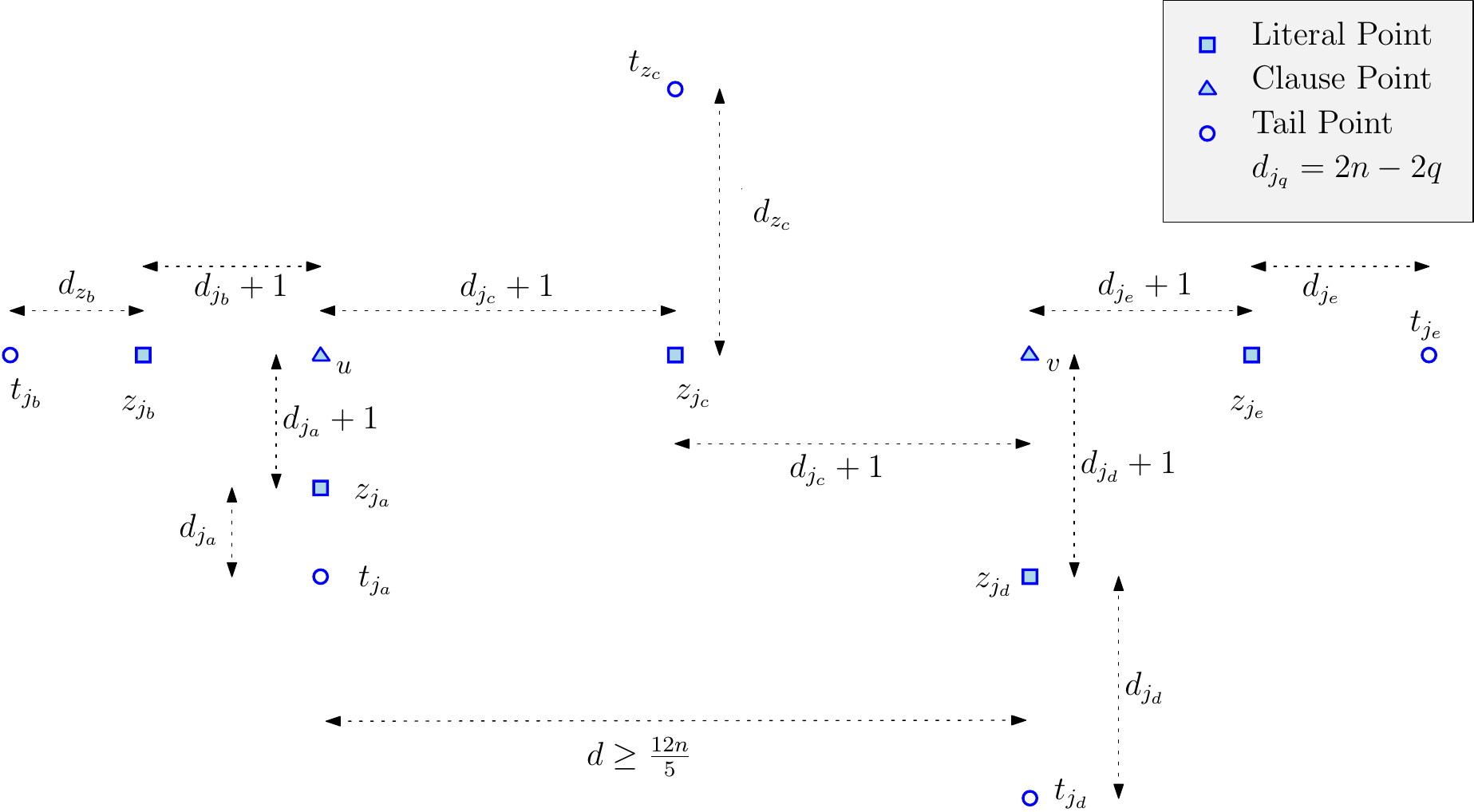}
    \caption{Illustration for the construction of the point set for a pair of intersecting clauses: $(j_a\vee j_b \vee j_c)$ and $(j_c\vee j_d \vee j_e)$. }
    \label{fig:cons}
\end{figure}

\noindent
\textbf{Construction of the Point Set:} 
The point set includes one point per clause, which is called a \emph{clause point} and one point per literal,  which is called  a \emph{literal point}. For each literal, we add a point, which we call a \emph{tail point} of that literal point. For example, consider a pair of clauses with $k$ distinct literals and assume that these clauses have a literal in common. The corresponding point set consists of two clause points, $k$ literal points and $k$ tail points.  
  Figure~\ref{fig:cons} gives an illustration with $k=5$. 
  
We now describe the construction in detail. We call a clause (or a pair of clauses that share a common literal) \emph{independent} if it does not intersect any other clause. We place these independent elements (clauses or pairs of clauses) far from each other such that points in one element are at least $n^2$ units apart from the points in any other element.

If a literal is shared among a pair of clauses, then we call it an \emph{intersection literal}. If each of the two clauses  corresponding to an intersection literal is of size three (i.e., contains three literals), then we call intersection literal   \emph{heavy} and otherwise, we call it \emph{light}.  Note that there can be at most $\frac{2n}{5}$ heavy literals (as each such literal corresponds to an independent pair of clauses with   $5$ distinct literals). We now relabel the variables (with labels $1,\ldots,{n}$) such that variables that correspond to heavy literals get  lower labels. For a variable with label $k$, we label its literals as $j_k$ and $\overline{j_{k}}$.  Thus for any heavy literal $j_k$,  we have $k \leq \frac{2n}{5}$.



We first describe the construction of the points corresponding to a heavy literal.  Let $(j_a\vee j_b \vee j_c)$ and $(j_c\vee j_d \vee j_e)$ be the corresponding clauses. For an integer $q$ let $d_{j_q}$ to be the  distance of $(2n-2q)$ units. In general, when  placing the  literal point $z_{j_q}$, we ensure that it is at distance $(d_{j_q}+1)$ from its corresponding clause point. When  placing a tail point $t_{j_q}$, we ensure that it is at a distance $d_{j_q}$ from its corresponding  literal point $z_{j_q}$. Therefore, in the following we only describe the position of the literal, clause and tail points relative to each other.

We place a clause point $u$ for $(j_a\vee j_b \vee j_c)$.  We place the literal points $z_{j_b}$ and $z_{j_c}$ to the left and right of the $u$ (on the horizontal line through $u$), respectively. We place the literal point $z_{j_a}$ vertically below $u$.  We then place  the clause point $v$ and its literal points to the right of the intersection literal $z_{j_c}$ symmetrically, as shown in Figure~\ref{fig:cons}. We place the tail points $t_{j_b}$ and $t_{j_e}$ to the left and right of $z_{j_b}$ and $z_{j_e}$, respectively. The tail points $t_{j_a}$ and $t_{j_d}$ are placed vertically below  $z_{j_a}$ and $z_{j_d}$, respectively. Finally, the tail point $t_{j_c}$  is  placed vertically above  $z_{j_c}$. 




For the light literals, we have at most 4 distinct literal points to place. The construction is the same as above and we intentionally put three literal points on the line passing through the clause points and the remaining one (if exists) below its corresponding clause point.  

For each independent clause, the construction is again the same as for $(j_a\vee j_b \vee j_c)$. If it contains two literals, then we place both literal points on the horizontal  line passing through the clause point.

\begin{remark}\label{rem:properties}
For a literal $z_q$, the construction ensures  the following properties. 
\begin{enumerate}
    \item [P$_1$:] The distance of $t_{z_q}$ is strictly greater than $2n-2q+1$ from all literal points except for its own literal point.
    \item [P$_2$:] The distance of  $z_{j_q}$ is strictly greater than $2n-2q+1$ from all tail points except for its own tail point.
\end{enumerate}
\end{remark}

Remark~\ref{rem:properties} is straightforward to verify from the construction except for the case when a heavy literal correspond to two literal  points below the line through the clause points. It may initially appear that if the construction places them in close proximity, then the nearest tail point of one literal point may be the tail point of the other literal point.  However, such a scenario does not appear due to our initial relabelling of the literals, i.e., the smallest distance between the vertical lines through these literal points is at least 
$d= 2(2n - 2(\frac{2n}{5})) = 2(\frac{6n}{5}) = \frac{12n}{5} > 2n$ (Figure~\ref{fig:cons}). 

\smallskip
\noindent
\textbf{Reduction:} First we show that if $I$ is satisfiable, then $P$ can be burnt in $2n$ steps using only the literal points as sources. For every $r$ from 1 to $n$, if $j_r$ is true then burn $z_{j_r}$ at the $(2r-1)$th step and burn $z_{\overline{j_{r}}}$ at the $2r$th step. If $j_r$ is false, then swap the steps. Any  tail point $t_{j_a}$  is at a distance of $(2n - 2a)$ from $z_{j_a}$. Even if $j_a$ is false, it still has $(2n - 2a)$ steps left to burn, which will be enough to burn the tail point. It only remains to show that all clause points are burned. Let $u$ be any clause point, as $I$ is satisfiable, at least one of its variable $j_r$ must be true, which will burn for at least $2n-(2r-1) = 2n-2r+1$ steps and thus burn $u$. 

We now assume that there is a burning sequence that burns all the points of $P$ using the literal points as sources. To construct a satisfying truth assignment for the LSAT, we  use the following lemma. 

\begin{lemma}\label{lem:app}
For a variable with label $r$, its literal points $z_{j_r}$ and $z_{\overline{j_{r}}}$ are burnt in the $(2r-1)$th and $2r$th step, respectively (or in the reverse order).
\end{lemma}

\label{app:hard}
\begin{proof} Consider the base case, i.e., the variable with label 1. The distance between $z_{j_1}$ (or, $z_{\overline{j_1}}$) and $t_{j_1}$ is $d_{j_1} = (2n-2)$. By Remark~\ref{rem:properties}, every other literal point is at least $(2n-1)$ distance apart from $t_{j_a}$. Therefore, $z_{j_1}$ and $z_{\overline{j_2}}$ must be burnt within the first two steps.


Assume now that the lemma statement holds for  the variables with labels $1,2,\ldots,(r-1)$. 
Consider now that tail point $t_{j_r}$. 
We will first show that for any $k<r$, the distance between $z_{j_k}$ and  $t_{j_r}$ is at least $(2n-2k+1)$ (Remark~\ref{rem:properties}), i.e.,  the fire at  $z_{j_k}$ cannot burn $t_{j_r}$. This can be verified  since the distance of $z_{j_k}$ from its clause point is $(2n-2k+1)$ and all other tail points are strictly further than this. Hence, it only remains to be shown that for any $k>r$, the distance between $z_{j_k}$ and  $t_{j_r}$ is at least $(2n-2r+1)$  (Remark~\ref{rem:properties}). This can be verified since the  distance of $z_{j_r}$ from its clause point is $(2n-2r+1)$ and all other literal points are strictly further than this. Hence, $t_{j_r}$  must only be burnt by the fire of $z_{j_r}$.  Consequently, $z_{j_r}$  must be burnt either at the $(2r-1)$th or $2r$th step. Similarly, we can show that $z_{\overline{j_{r}}}$ must be burnt either at step $(2r-1)$ or $2r$. 
\end{proof}

We now construct a satisfying truth assignment as follows. If $z_{j_r}$ or $z_{\overline{j_{r}}}$ is burnt in odd step, we set it to be true. Otherwise, we set it to be false. We now show that $I$ is satisfied. Assume for a contradiction that there is a clause $c$ which is not satisfied, i.e., all of its  associated literal points are burnt in even steps. By Lemma~\ref{lem:app}, each literal point $z_{j_a}$ corresponding to $c$ burns for  $(2n-2a)$ steps, which is not enough to burn the clause point. This contradicts our initial  assumption of a valid burning sequence. This completes the NP-hardness reduction.


\begin{theorem}
Given a set  of points $P$ in the plane  and a  subset $S$ of $P$, it is NP-hard to construct a point burning sequence using only the points of $S$ that burns all the points of $P$  within $|S|$ rounds. 
\end{theorem}

\subsection{Burning Maximum Number of Points}

Our NP-hardness result implies that given a  subset $S=\{s^1,\ldots,s^q\}$ of $q$ points from $P$, it is NP-hard to burn the maximum number of points  by only burning the points of $S$ within $q$ rounds. We now show how to obtain a 2-approximation for this problem. 
For every point $s^j$, where $1\le j\le q$, we consider $q$ sets. The $i$th set $\Delta^j_i$, where $1\le i\le q$,  contains the points that are covered by the disk of radius $i$ centered at $p$, i.e., these points are within a distance of $i$  from $p$. We thus have a collection of sets $\{\Delta^1_1,\ldots, \Delta^1_q, \ldots, \Delta^q_1,\ldots, \Delta^q_q\}$, which can be partitioned into 
$q$ groups based on radius, i.e., the $i$th group contains the sets $\{\Delta^1_i,\ldots,  \Delta^q_i\}$. 
 To burn the maximum number of points by burning $S$, we need to select one subset from each radius group so that the cardinality of the union of these sets is maximized. This is exactly the maximum set cover problem with group budget constraints, which is known to be 2-approximable~\cite{chekuri2004maximum}.

 \begin{theorem}
Given a set $P$ of $n$ points in the plane and a subset $S$ of $q$ points from $P$, one can compute a point  burning sequence using $S$ within $q$ rounds in polynomial time that burns at least half of the maximum number of points that can be burned using $S$ within $q$ rounds. 
 \end{theorem}
 
 If we want to burn maximum number of points within $q$ rounds, then we can set $S$ to be equal to $P$ to have a 2-approximate solution. We thus have the following corollary.
 
 \begin{corollary}
Given a set of $n$ points in the plane and a positive integer $q<n$, one can compute a point  burning sequence in polynomial time that burns at least half of the maximum number of points that can be burned within $q$ rounds. 
 \end{corollary}

\section{Conclusion}

In this paper we have shown that point burning and anywhere burning problems admit PTAS in one dimension and improved the known approximation factors in two dimensions. To improve the previously known approximation factor in two dimensions we used a geometric covering argument. We believe our covering strategy can be refined further by using a  tedious case analysis. However, this would not provide a PTAS. Therefore, the most intriguing question in this context is whether these problems admit PTAS in two dimensions.

We have also proven that the problem of burning the maximum number of points within a given number of rounds is NP-hard, but 2-approximable by a known result on set cover with group budget constraints. It would be interesting to design a better approximation algorithm  leveraging the geometric  structure. 



 


\bibliographystyle{splncs04}
\bibliography{biblio} 
 
\end{document}